\documentclass[%
 aip,
 amsmath,amssymb,
 longbibliography
 reprint,%
12pt,
onecolumn
]{revtex4-1}

\usepackage{graphicx}% Include figure files
\usepackage{dcolumn}% Align table columns on decimal point
\usepackage{bm}% bold math
\usepackage[utf8]{inputenc}
\usepackage[T1]{fontenc}
\usepackage{mathptmx}
\usepackage{etoolbox}
\usepackage{bm}% bold math
\usepackage{setspace}
\usepackage{amsmath}
\usepackage{amssymb}
\usepackage{graphicx}
\usepackage{placeins}
\usepackage{setspace}
\usepackage{fullpage}
\usepackage{caption}
\usepackage{anyfontsize}
\usepackage[colorlinks=true,linkcolor=blue]{hyperref}
\hypersetup{
    colorlinks=true,
    linkcolor=blue,
    citecolor=blue,
    filecolor=magenta,      
    urlcolor=blue,
}
\makeatletter
\def\@email#1#2{%
 \endgroup
 \patchcmd{\titleblock@produce}
  {\frontmatter@RRAPformat}
  {\frontmatter@RRAPformat{\produce@RRAP{*#1\href{mailto:#2}{#2}}}\frontmatter@RRAPformat}
  {}{}
}%
\makeatother
\begin{document}

%\preprint{AIP/123-QED}

\title[]{Metamorphosis of transition between states of limit cycle oscillations in  aeroacoustic system}
% Force line breaks with \\
\author{Siddharth Kancharlapalli}
\author{ Beeraiah Thonti}
%\author{R. I. Sujith}

\author{Sivakumar Sudarshanan} 
\author{Ramesh S. Bhavi } 
\author{R. I. Sujith}

\affiliation{Department of Aerospace Engineering, Indian Institute of Technology Madras, Chennai, Tamil Nadu 600036, India.}
\affiliation{Centre of Excellence for Studying Critical Transitions in Complex Systems, Indian Institute of Technology Madras, Chennai, Tamil Nadu 600036, India.}
\date{\today}% It is always \today, today,
             %  but any date may be explicitly specified

\maketitle
\doublespacing
\section{Abstract}
 Dynamical systems undergoing transition to oscillatory state exhibit change in the nature of the transition from supercritical to subcritical Hopf bifurcation or vice versa upon variation of a secondary parameter. This phenomenon is referred to as change of criticality. Many real-world systems undergo transition to oscillatory state that do not fit in the framework of Hopf bifurcation, and hence the change of criticality. We perform experiments on a ducted turbulent aeroacoustic flow constrained by two orifices separated at a distance apart. We vary the Reynolds number ($Re$), a bifurcation parameter causing a transition between various limit cycles. We change the distance between the orifices as the secondary parameter. We discover that turbulent aeroacoustic flows exhibit a metamorphosis of the transition from continuous to abrupt through a canard explosion, a bifurcation unique for its continuous yet rapid nature.  We observe two distinct abrupt bifurcations, differing in their dynamical states associated with the transition. Understanding this metamorphosis from continuous to abrupt aids in developing low-cost control and preventive strategies for systems undergoing a route to oscillatory instabilities.
 \section{Lead Paragraph}
Upon a change in a secondary parameter, dynamical systems exhibiting Hopf bifurcation undergo a change in the nature of the bifurcation from subcritical to supercritical or vice versa. This phenomenon is referred to as change of criticality. Real-world systems, for instance, complex turbulent reactive flows, exhibit a metamorphosis of the transition from continuous to abrupt. This metamorphosis delineates an intricate change in the nature of transition, which can't be explained within the framework of the change of criticality. Aeroacoustic flows are turbulent non-reactive flows, undergoing transition to high-amplitude acoustic pressure oscillations, referred to as aeroacoustic instability. Aeroacoustic instability is a manifestation of nonlinear interactions across scales between and within
subsystems (acoustic and hydrodynamic fields). These interactions, along with the inherent turbulence of the flow, give rise to complex and intriguing dynamics. We study the change in the nature of transition between states of aeroacoustic instability in a prototypical experimental setup consisting of ducted flow past two orifices. Through nonlinear time-series analysis, such as amplitude spectra and phase-space reconstruction, we identify different dynamical states associated during the transition. Our study reveals that turbulent flows undergo metamorphosis of transition, warranting a framework, beyond the change of criticality, to understand and compare different complex systems.

%\doublespacing
\section{Introduction}
Transitions between states of a system are ubiquitous; for example, systems spanning ecology, financial markets, society etc. exhibit transitions between their disparate states \citep{scheffer2009critical}. Often, these transitions are undesirable and understanding them helps us to develop better control strategies and early warning indicators \citep{Pavithran2021, sujith2021early, scheffer2009early}.  For example, the transition to oscillatory instability in fluid mechanical systems such as aeroacoustic systems is detrimental to many engineering configurations such as solid and hybrid rocket motors \cite{brown1981vortex,dunlap1981exploratory,mettenleiter2000adaptive}, ramjet engines \cite{jou1990modes}, industrial pipeline systems\cite{ziada2010flow}, cavities in gas dynamics lasers, supersonic intakes, steam regulation, control valves and hydraulic gates \cite{doi:10.2514/3.8130}. These oscillatory instabilities, referred to as aeroacoustic instability, manifest as high-amplitude periodic oscillations. Such high-amplitude acoustic pressure oscillations result in noise pollution, induce fatigue stress, and jeopardize the structural integrity of the system.\cite{ziada2014flow}. In contrast, the transition to aeroacoustic instability of wind instruments such as flute and flue organ pipe produces a pleasant sound \cite{fabre2012aeroacoustics}.
\par Based on the classification suggested by \citet{10.1115/1.3448624}, all the foregoing examples of aeroacoustic sound production either fall under the category of fluid-resonant or fluid-dynamic cavity flows. The investigations of this study belong to the latter group, in which the transition to the aeroacoustic instability happens due to the feedback between the hydrodynamic and acoustic fields.  This feedback could be initiated by flows through various duct-orifice/cavity geometries \cite{10.1115/1.3448624}. One of these is the flow through a duct obstructed by two orifices placed at a distance apart. During the state of aeroacoustic instability, the frequency of self-sustained large amplitude acoustic pressure oscillations is close to the natural harmonics of the duct. Moreover, previous experimental studies have confirmed the excitation of a particular harmonic of the duct more than once as the velocity of the fluid through the duct is increased systematically \cite{aly2010flow,nomoto1982experimental}. This is attributed to the coherent structure of the shear layer in between the orifices \cite{chung1986interactions,huang1991active}. 
\par As the flow passes through the first orifice, it develops into a shear layer. During the state of aeroacoustic instability, the vortices shed within the shear layer develop into coherent structures downstream, before impinging onto the second orifice \cite{nomoto1982experimental,karthik2008mechanism}. When this impingement of the coherent structures on the downstream orifice at a particular velocity of the flow is in phase with the acoustic perturbations in the duct, the flow-structure interaction leads to the production of acoustic power sources (power sources for the acoustic field in the duct). Various numerical and experimental studies \cite{geveci2003imaging,tan2003sources}, based on Howe's theory of aeroacoustic sound production \cite{howe2001vorticity},  have obtained spatial acoustic power field in different duct-orifice/cavity geometries. Furthermore, the acoustic velocity perturbations at the sensitive region of the shear layer, i.e., near the first orifice \cite{doi:10.2514/3.8130} synchronize the vortex shedding phenomenon with the acoustic field within the duct \cite{blevins1985effect,hong2020frequency}. When the convective velocity of these shedded vortices is in such a way that the impingement of vortices on the second orifice is in phase with the acoustic perturbations, the condition for the positive feedback between the two fields is set up \cite{huang1991active,hourigan1990aerodynamic}. \citet{flandro1986vortex} has analytically studied this feedback by considering the effect of orifices on the flow through a pseudosound field (non-propagating pressure field), restricted to the region in between the orifices, due to the impingement of the shear layer. In another notable analytical study,
 \citet{chung1986interactions} models the feedback by direct superposition of the acoustic and velocity fields in the linearised governing equations. 
\par Owing to the turbulence in the flow field, aeroacoustic flows exhibit dynamics characterized by multiple scales. The interactions within and across the two subsystems, hydrodynamic and acoustic fields, at different length scales, leads to the emergence of periodic oscillations (order) from aperiodic fluctuations \cite{Pavithran_2020,nair2016precursors}. The aforementioned interactions, along with the inherent turbulence, warrant the aeroacoustic flows to be a complex system, exhibiting diverse dynamical behavior \cite{bhavi2024dynamical,10.1115/GT2024-127399}. \citet{matveev2005reduced} had modeled the nonlinear feedback between the subsystems as a kicked forcing to the acoustic pressure oscillator, assuming the vortex impingement to be instantaneous and spatially localized. %Furthermore, using the same model,  Singh.et.al \cite{singh2024reduced} has studied various dynamical states observed in the transition to aeroacoustic instability. 
In addition, \citet{10.1115/GT2024-127399} has modeled the nonlinearities during the transition to instability using the Rayleigh-Duffin Van der Pol oscillator. Previous studies have investigated the transition to aeroacoustic instability under the lens of nonlinear dynamical systems theory \cite{Pavithran_2020,boujo2020processing}. The dynamical state during the state of self-sustained periodic oscillations is limit cycle oscillations (LCO) \cite{nair2016precursors}. During the stable state of the flow, the acoustic pressure fluctuations exhibit low amplitude aperiodic fluctuations \cite{bhavi2024dynamical}.
\par In nonlinear dynamical systems theory, bifurcation is the change in the qualitative behavior of the system upon varying the control parameter beyond a critical point \cite{strogatz2001nonlinear}. This control parameter is referred to as the bifurcation parameter or the primary parameter. The bifurcation of aeroacoustic flows from the low amplitude aperiodic state to the state of large amplitude LCO has previously been studied with the velocity of the flow through the duct as the bifurcation parameter \cite{bhavi2024dynamical, 10.1115/GT2024-127399}. Furthermore, prior to the onset of these self-sustained oscillations, aeroacoustic flows exhibit a state of low amplitude aperiodic fluctuations amid bursts of periodic oscillations in the unsteady acoustic pressure fluctuations \cite{nair2016precursors,bhavi2024dynamical}. This state of intermittency is also observed in turbulent reactive flow systems \cite{nair2014intermittency}. Furthermore, it is worth mentioning that \citet{bourquard2021whistling} have proposed necessary conditions for a state of intermittency by considering nonlinear reactive and resistive coupling between two linear oscillators (acoustic and hydrodynamic fields) constructed from the measurements of acoustic admittance of the cavity. Further, \citet{bhavi2024dynamical} has shown vortex acoustic lock-in using the theory of synchronization. Moreover, \citet{10.1115/GT2024-127399} have constructed an Arnold tongue for lock-in between acoustic and hydrodynamic fields both experimentally and from models.
\par Moreover, dynamical systems, upon a change in one of their additional parameters, exhibit a change in the nature of the bifurcation. For instance, in a Rijke tube (a prototypical laminar thermoacoustic system), as the power of the heating mesh (bifurcation parameter) is increased, the quiescent (fixed point) operating state loses its stability and the system exhibits limit cycle oscillations (Hopf bifurcation) \cite{10.1063/5.0091826}. \citet{etikyala2017change} have shown that at lower mass flow rates through the tube, this bifurcation is supercritical, i.e., a continuous bifurcation. However, when the mass flow rate through the tube (secondary parameter of the system) is increased to a higher value, the study showed that the system undergoes a subcritical Hopf bifurcation, i.e., an abrupt bifurcation from a quiescent state to a state of self-sustained oscillations. Such abrupt transitions have received considerable interest in the literature owing to their detrimental consequences in many applications. Many other systems, such as ecology \cite{bestelmeyer2011analysis}, brain \cite{durstewitz2010abrupt}, climate \cite{lockwood2001abrupt} and turbulent reactive flow systems \cite{PAWAR20216193,bhavi2023abrupt} exhibit abrupt transitions. Similar to the laminar thermoacoustic system (Rijke tube), due to the change in the governing nonlinearities upon secondary parameter variation, many systems undergo change in the nature of the bifurcation from continuous to abrupt \cite{rose1981ecological,zhang2008polarization,verma2013supercriticality}. For systems undergoing the route of oscillatory instabilities, this change in the nature of the transition from smooth to abrupt is shown to be a universal phenomenon \cite{doi:10.1126/sciadv.abe3824}. 
\par However, a recent experimental study in a complex, turbulent reactive flow system showed that the nature of the transition from a low amplitude chaotic state to high amplitude instability undergoes an intricate change from continuous to abrupt \cite{qn17-x37z}. The authors obtained this transition by varying the equivalence ratio of the fuel-oxidizer mixture entering the combustor as the bifurcation parameter. They have discovered five different routes of this bifurcation by changing the bluff body position in the combustor and the thermal power input as the secondary parameter individually. This intricate change is attributed to the inherent turbulence in the system. Turbulence is a complex, multifractal phenomenon \cite{mukherjee2024turbulent}. As a result, the intricacy of the nonlinear interactions between subsystems at different scales is bound to increase. In line with this, the classic notion of change from continuous to abrupt (or supercritical to subcritical) fails to describe the change in the nature of bifurcation in turbulent systems.  These findings suggest that turbulence (or complexity) entails intricate change in the nature of transition, beyond change of criticality, in systems undergoing the route to oscillatory instabilities. Modeling dynamics based on this intricate phenomenology would further enhance the understanding of the governing nonlinearities in the system. These insights could be used to develop effective control and mitigation strategies. 
\par By varying the velocity through the duct as a bifurcation parameter, \citet{10.1115/GT2024-127399} have shown that bifurcation from the low amplitude state to the state of instability happens through a state of quasi-periodic oscillations followed by a state of period-2 oscillations. In line with the fact that hydrodynamic and acoustic fields could couple at multiple frequencies \cite{chung1986interactions}, as the velocity of the flow through the duct is increased, previous experimental studies have reported a shift in the frequency of the self-sustained oscillations \cite{sano2008transition,karthik2008mechanism,testud2009whistling}. With the velocity through the duct as the bifurcation parameter, \citet{bhavi2024dynamical} have shown that this shift in the frequency could either occur directly, via aperiodic oscillations or via a state of intermittency. However, the effect of change in the distance between the orifices, i.e., the distance between the point of separation and the point of impingement of the shear layer, on the nature of the bifurcation between LCOs, especially the intricacies caused by the inherent turbulence in the flow, is not fully understood.  
\par To this end, we conduct experiments in a prototypical aeroacoustic system. We vary the Reynolds number of the flow ($Re$) in a quasi-static manner and observe bifurcation between LCOs. At each $Re$, we obtain the acoustic pressure fluctuations downstream of the second orifice. Using nonlinear time series analysis, we characterize the different dynamical states associated with the bifurcation between the LCOs. Further, we investigate the change in the nature of this bifurcation by changing the spacing between the orifices as a secondary parameter. We discover that the nature of the transition between LCOs undergoes metamorphosis from continuous to abrupt through a canard explosion. Moreover, we observe a change in the sequence of dynamical states between the LCOs associated with this metamorphosis.

\section{Experimental setup}

\begin{figure*}[h!]
\centering
\includegraphics[]{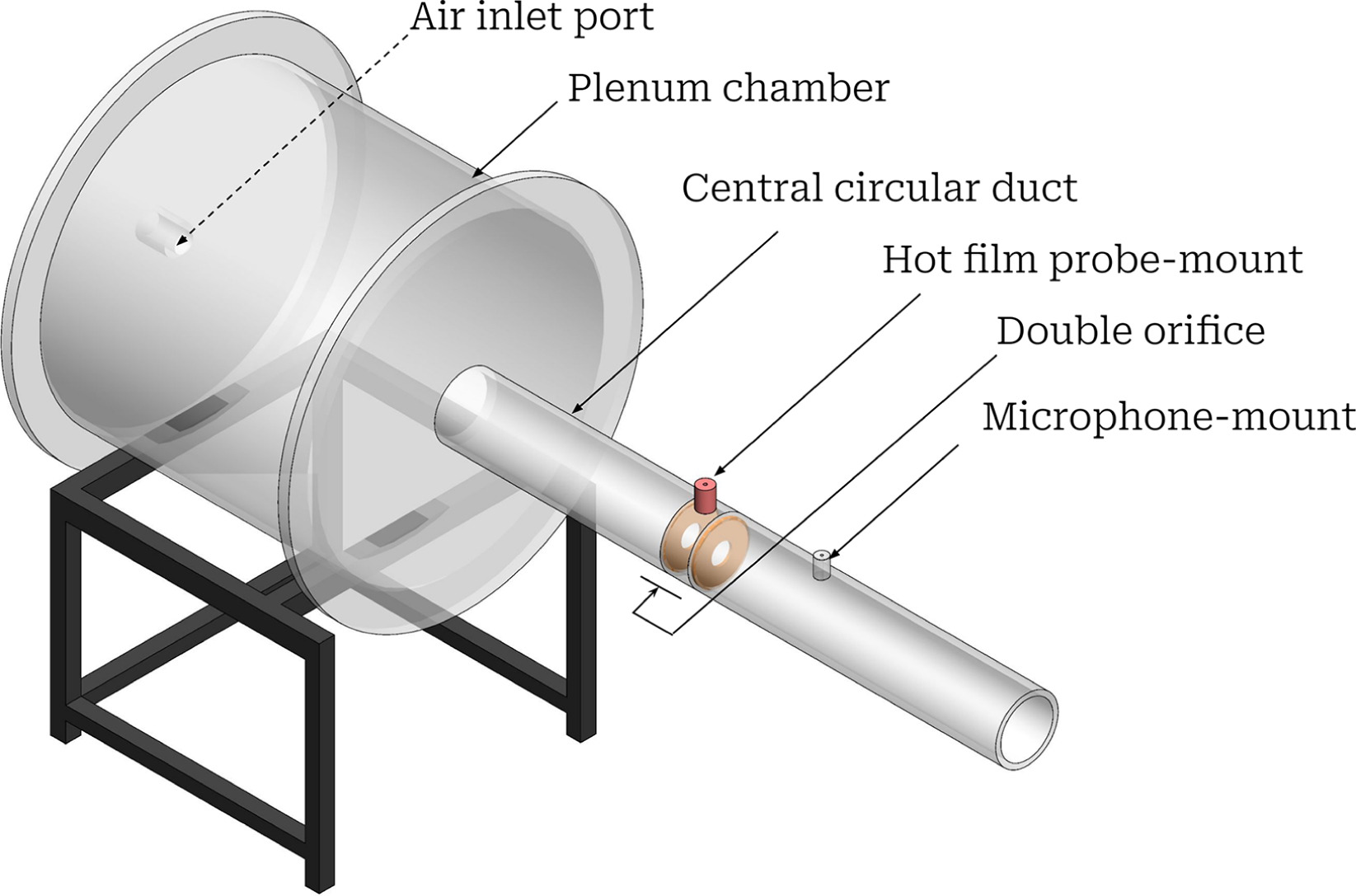}
\captionsetup{justification=justified, format=plain, singlelinecheck=false}
\caption{Schematic of the experimental setup consisting of a ducted flow through two orifices separated at a distance apart.
}
\label{setup}
\end{figure*}

The experimental setup employed in this study consists of two orifices located inside a circular housing. Firstly, the air enters the plenum chamber through an inlet port. The plenum chamber is used to minimize unsteady fluctuations in the incoming flow. Following this, the flow enters a circular duct (of length 280 mm and inner diameter of 55 mm) mounted to the plenum chamber. At the end of the circular duct, a circular housing (a duct of length 80 mm) having the same inner diameter as the outer diameter of the circular duct is attached to facilitate the accommodation of two orifices (refer Fig~\ref{setup}). The separation between the orifices (16 mm orifice diameter and 2.5 mm thickness) is varied as a secondary control parameter for this study. Another circular duct (of length 300 mm and radial dimensions same as the circular duct mounted to the plenum chamber) is located within the circular housing, at the end of the second orifice (located downstream to the flow).  On the walls of the circular duct, at distance of 100 mm downstream of the second orifice, a microphone is mounted to acquire the acoustic pressure fluctuations.

The experiments were conducted at room temperature. The Reynolds number ($Re$) based on the orifice spacing is varied as a control parameter for this study. $Re$ is computed using the formula $Re=\rho \bar{u} d/\mu$, where $\rho$ is the density (kg/m$^3$), $\bar{u}$ is the mean velocity (m/s), $d$ is the spacing between the orifices (m), and $\mu$ is the viscosity of air at room temperature with units of Pa·s (Pascal-seconds). We vary the $Re$ through two Alicat (MCR series) mass flow controllers (MFCs) in a parallel configuration with a measurement uncertainty of $\pm$ (0.8\% of reading and 0.2\% of full scale reading). One of the MFCs has a range of 500 SLPM (standard liters per minute), while the other has a range of 10 SLPM. With the help of 10 SLPM MFC, we varied $Re$ in fine steps (0.5 SLPM during the continuous and abrupt bifurcation and 0.25 SLPM during the canard explosion) during the onset of the bifurcation between LCOs.

The spacing between the orifices is decreased from $26 \, \text{mm}$ to $10 \, \text{mm}$ in steps of $2 \, \text{mm}$. For every distinct orifice spacing, $Re$ is varied in a quasistatic manner to study the transition between LCOs. The range of $Re$ corresponding to the transition between desired LCOs is different for different orifice spacings. When the orifices are spaced from $24 \, \text{mm}$ to $10 \, \text{mm}$, the required maximum and the minimum mass flow rates are $60 \, \text{SLPM}$ and $252 \, \text{SLPM}$ respectively. This variation is achieved by an increment of $3 \, \text{SLPM}$ through the MFC having a range of $500 \, \text{SLPM}$. For example, this corresponds to the variation in $Re$ from $2900 \pm 150$ to $29000 \pm 350$ when the orifices are spaced at $24 \, \text{mm}$ apart. 

When the orifices are spaced at $26 \, \text{mm}$ apart, the mass flow rates through the duct are varied from $80 \, \text{SLPM}$ to $302 \, \text{SLPM}$ in steps of $4 \, \text{SLPM}$. This corresponds to the variation in $Re$ from $10000 \pm 500$ to $37000 \pm 500$. Note that, whenever necessary, MFC having a range of $10 \, \text{SLPM}$ is used in parallel with $500 \, \text{SLPM}$ MFC to study the behavior of the system upon fine increments in $Re$. For example, when the orifices are spaced at $24 \, \text{mm}$ apart, a MFC having a range of $10 \, \text{SLPM}$ is required to increase the mass flow rates through the duct from $122 \, \text{SLPM}$ to $132 \, \text{SLPM}$ in steps of $0.5 \, \text{SLPM}$. This corresponds to the variation in $Re$ from $14000 \pm 0.27$ to $15000 \pm 11$.

The pressure field pre-polarised microphone (Piezoelectronics PCB378C10) mounted downstream of the second orifice, through which we acquire the acoustic pressure fluctuations, is equipped with a preamplifier system and a condenser. The data is acquired for $5 \, \text{s}$ at a sampling frequency of $10 \, \text{kHz}$.  The measurements from the microphone has an inherent noise of $7 \, \text{mPa}$.

\section{Results}

% \begin{figure} [hbt]
%     \centering
%     \includegraphics[width=1\linewidth]{transition1.png}
%     \caption{Change in the nature of the transition between states of limit cycle oscillations from continuous to abrupt via a canard explosion, upon decrease in the spacing between the orifices. $Re$ is varied as the control parameter for each distinct orifice spacing ($d$ mm). For the clarity of representation, the variation of $p'_{\mathrm{rms}}$ with $Re$ is represented as a function of $\tfrac{\mathrm{Re} \times 10^{-2}}{d}\ (\mathrm{mm}^{-1})$ for every distinct orifice spacing ($d$ (mm)). Transitions represented by purple are continuous, by magenta are characterized by a canard explosion, and by orange and by yellow are abrupt. The abrupt bifurcations observed upon an increase in the $Re$ are marked by an arrow.}
%     \label{Global}
% \end{figure}

\begin{figure*}[h!]
\centering
\includegraphics[height=10cm]{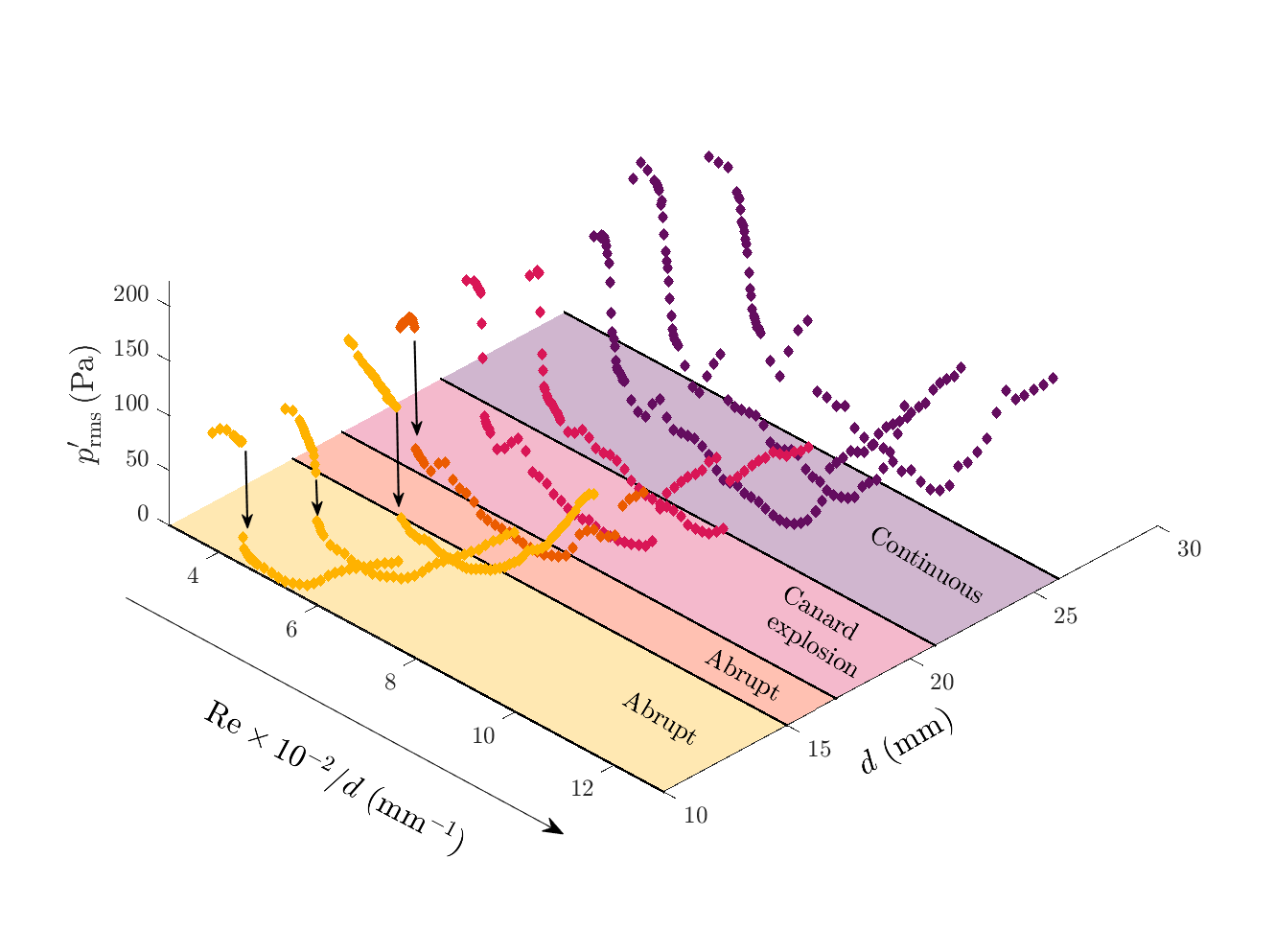}
\captionsetup{
  justification=raggedright,
  singlelinecheck=false
}
\caption{Change in the nature of the transition between states of limit cycle oscillations from continuous to abrupt via a canard explosion, upon decreasing the spacing between the orifices. $Re$ is varied as the control parameter for each distinct orifice spacing ($d$ mm). For the clarity of representation, the variation of $p'_{\mathrm{rms}}$ with $Re$ is represented as a function of $\mathrm{Re}\times10^{-2}/d\;(\mathrm{mm}^{-1})$ for every distinct orifice spacing ($d$ (mm)). Transitions represented by purple are continuous, by magenta are characterized by a canard explosion, and by orange and by yellow are abrupt. The abrupt bifurcations observed upon an increase in the $Re$ are marked by an arrow.}
\label{Global}
\end{figure*}

Figure~\ref{Global} represents the variation of $p'_{\mathrm{rms}}$ with $Re$ for different orifice spacing. For every orifice spacing, the $Re$ is varied to obtain a transition between limit cycle oscillations (LCO). The dynamical state during the state of aeroacoustic instability is an LCO.  The feedback between acoustic and hydrodynamic subsystems enables an aeroacoustic system to be excited at different frequencies \cite{bhavi2024dynamical,chung1986interactions}. Often, these frequencies correspond to the natural frequencies of the duct \cite{aly2010flow,nomoto1982experimental}. For every orifice spacing ($d$ mm), $Re$ is varied to study the transition between LCOs. That is, the dynamical states at the beginning and the end of every transition is a LCO of the same frequency. However, from the perspective of nonlinear dynamics, the nature of these two LCOs is different, which we will discuss in detail in further sections. Instead of just $Re$, for the clarity of representation, we plot the variation of  $p'_{\mathrm{rms}}$ with $\mathrm{Re}/d$ ($Re$ normalized with the orifice spacing) for different orifice spacings.  We observe a change in the nature of the transition from continuous (transitions represented by purple in Fig~\ref{Global}) to abrupt (transitions represented by orange and yellow) via a canard explosion (transitions represented by magenta). Associated with this change in the nature of the transition, we observe a change in the sequence of dynamical states associated with the transition between LCOs. Upon backward variation of the control parameter, while abrupt bifurcations exhibit hysteresis, continuous bifurcations and canard explosions retrace their path. Moreover, canard explosions exhibit sensitivity to the fluctuations in the control parameter during the bifurcation.

\subsection{Continuous transition}
\begin{figure*}
\centering
\includegraphics[height=16cm]{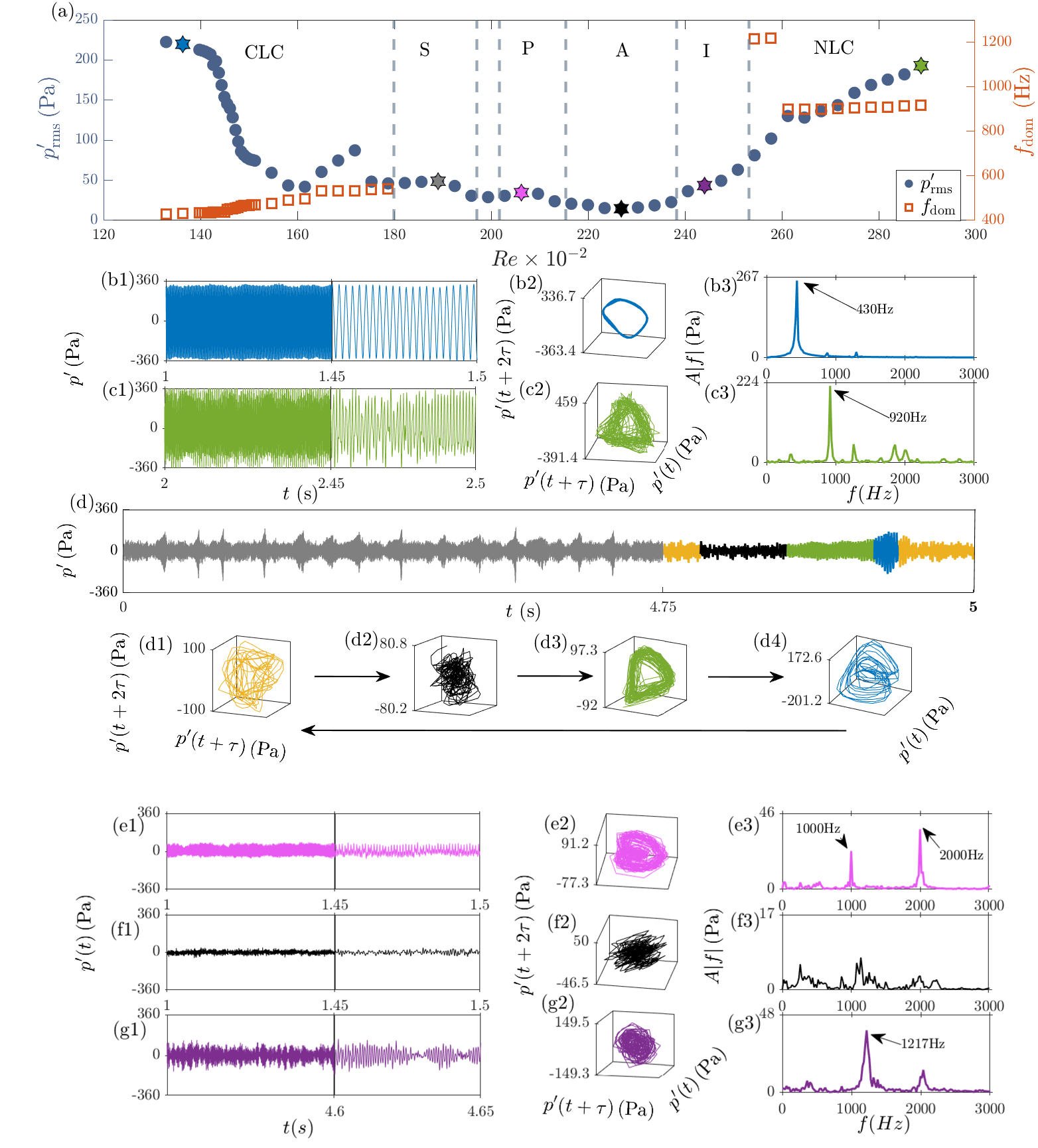}
\captionsetup{
  justification=raggedright,
  singlelinecheck=false
}
\caption{
Continuous transition between states of limit cycle oscillations. (a) Variation of $p'_{\mathrm{rms}}$ and $f_{\mathrm{dom}}$ with $Re$ during the transition when the orifices are spaced at 24 mm apart. The bifurcation from clean limit cycle (CLC) to noisy limit cycle (NLC) follows the sequence: transient regime characterized by brief high amplitude period 2 epochs amidst NLC, period 2 oscillations, aperiodicity, and intermittency. While (b1),(b2) and (b3) represent the $p'$, reconstructed phase space and the amplitude spectrum during the state of CLC, (c1),(c2) and (c3) correspond to the state of NLC. (d) Represents the acoustic pressure fluctuations ($p'$) corresponding to the transient regime. The evolution of the windowed phase space reconstruction unravels the switching between NLC (yellow epochs) and brief high amplitude period 2 oscillations (blue epochs) (d1),(d2),(d3), and (d4). We represent the $p'$, reconstructed phase space and the amplitude spectrum of the $p'$ for the rest of the states. Namely, (b1-b3) correspond to CLC, (c1-c3) correspond to NLC, (e1-e3) correspond to period 2 oscillations, (f1-f3) correspond to aperiodic flutuations and (g1-g3) correspond to the state of intermittency.
}
\label{24mm_1}
\end{figure*}
Figure~\ref{24mm_1}(a) shows the variation of $p'_{\mathrm{rms}}$  with $Re$ and the frequency of oscillations associated with the states of limit cycle oscillations (LCO) ($f_{\mathrm{dom}}$) when the orifices are spaced at 24 mm apart. Upon increasing $Re$ systematically from 13000 to 30000 (refer to Fig~\ref{24mm_1}a), the system bifurcates from clean limit cycle oscillations (states identified as CLC in Fig~\ref{24mm_1}a) to noisy limit cycle oscillations (states identified as NLC in Fig~\ref{24mm_1}a). Firstly, for a range of $Re$ values from 13000 to 18000, we observe that $p'$ exhibits CLCs. Figure~\ref{24mm_1}(b1) shows the time series of $p'$ corresponding to a value of $Re$ (marked by blue in Fig~\ref{24mm_1}a) belonging to the states of CLC. The clean limit cycle oscillations of the $p'$ is manifested as a very thin ring topology in its reconstructed phase space (Fig~\ref{24mm_1}b2). The corresponding amplitude spectrum exhibits a distinct peak at 430 Hz (Fig~\ref{24mm_1}b3).  Moreover, the frequency associated with these clean limit cycle oscillations, $f_{\mathrm{dom}}$ varies within this region of $Re$ in a non-trivial manner (refer Fig~\ref{24mm_1}a).
\par Aeroacoustic instability manifests as limit cycle oscillations in $p'$. During the state of aeroacoustic instability, the vortex shedding near the first orifice is in phase synchrony with $p'$ within the duct. Within the region of $Re$ corresponding to the states of CLC, when the value of $Re$ is increased beyond 14000, we observe a gradual decline of $p'_{\mathrm{rms}}$ (refer to Fig~\ref{24mm_1}a). Associated with this gradual decline, $f_{\mathrm{dom}}$ shows an increasing trend with $Re$ from 430 Hz to 470 Hz. However, upon further increase in $Re$ beyond 16000, we observe an increasing trend of $p'_{\mathrm{rms}}$. Associated with this switch from a gradual declining to an increasing trend of  $p'_{\mathrm{rms}}$, the $f_{\mathrm{dom}}$ of the states of CLC switches from 470 Hz to 530 Hz and remains at 530 Hz for the rest of the region.
\par As $Re$ is further increased beyond the state of CLC, we observe that the system bifurcates to the state of NLC through various dynamical states. During the state of NLC, the periodic oscillations of $p'$ is contaminated with noise. Figure~\ref{24mm_1}(c1) shows the time series of $p'$ of one of such states exhibiting noisy periodic oscillations. The deviation from the clean limit cycle behavior is evident in the irregular modulations of the amplitude of $p'$ (refer Fig~\ref{24mm_1}b1 and c1). This manifests as a thicker ring topology in the reconstructed phase space (Fig~\ref{24mm_1}c2). The thickness here is an indicator of the irregularity of the amplitude modulations or in other words, an indicator of NLC \cite{qn17-x37z}.  These noisy limit cycle oscillations exhibit a peak at 940 Hz in the corresponding amplitude spectrum (Figure ~\ref{24mm_1}c3). However, at the value of $Re$ of nearly 26000, the $f_{\mathrm{dom}}$ corresponding to the states of NLC shifts from 1220 Hz to 940 Hz (refer Fig~\ref{24mm_1}a). The bifurcation from the state of CLC, characterized by a range of frequencies, to the noisy periodic oscillations (NLC) happens through a sequence of dynamical states (states marked by S, P, A and I in Fig~\ref{24mm_1}a).
\par The first dynamical state associated with the bifurcation from CLC-NLC (refer state S in Fig~\ref{24mm_1}a) consists of signatures of LCOs corresponding to the states identified as CLC or NLC. We observe that $p'$ corresponding to this state, amidst noisy periodic oscillations, exhibits high amplitude epochs. The frequency of these noisy periodic oscillations (highlighted by yellow in Fig~\ref{24mm_1}d) corresponds to the states of CLC in the transition, post the frequency shift (i.e., the states of CLC post the shift in the trend of $p'_{\mathrm{rms}}$). The brief high amplitude epochs (highlighted with blue in Fig~\ref{24mm_1}d) are characterized as period 2 oscillations. The frequency associated with these oscillations is the first and the second harmonics of the frequency of oscillations associated with the states of CLC in the transition, before the gradual declining of $p'_{\mathrm{rms}}$. 
\par During the switching from the epochs of NLC to the high amplitude epochs, the increase in the amplitude of $p'$ is gradual; however, during the switching back to the epochs of NLC, the decrease in the amplitude of $p'$ is relatively steeper (refer to Fig~\ref{24mm_1}d, i.e., the variation of $p'$  during the shift from yellow to blue epochs is gradual and increasing, but from blue to yellow is steeper and decreasing). We observe that the switching from NLC (marked by yellow in Fig~\ref{24mm_1}d) to brief epochs of high amplitude period 2 oscillations (marked by blue in Fig~\ref{24mm_1}d) happens via two states. Firstly, $p'$ exhibits low amplitude aperiodic fluctuations (marked by black in Fig~\ref{24mm_1}d). Following this, during the gradual increase in the amplitude of $p'$  (marked by green in Fig~\ref{24mm_1}d), associated with the switching from NLC to the high amplitude epochs, the system exhibits NLC. The frequency associated with these green epochs corresponds to the frequency of the states identified as NLC in the transition, after the frequency switching (i.e., the states of NLC in the range of $Re$ of 26000 - 29000). In contrast to the switching from NLC (yellow epochs) to high amplitude period 2 epochs via aperiodicity followed by another NLC (green epochs), the system directly switches from high amplitude period 2 oscillations back to NLC (yellow epochs), evident in the relatively steep decline in the amplitude of $p'$ (refer Fig~\ref{24mm_1}d)
\par Figures~\ref{24mm_1}(d1), (d2), (d3) and (d4) shows the evolution of the reconstructed phase space during the switching to high amplitude period 2 epochs. Firstly, the system is characterized by a thicker ring topology (Fig~\ref{24mm_1}d1), followed by a clustered topology (Fig~\ref{24mm_1}d2), marking the transition from NLC to aperiodic fluctuations. Following this, the system switches to another thicker ring topology (Fig~\ref{24mm_1}d3). Finally, the system switches to high amplitude period 2 epochs. The phase space reconstruction exhibits a twisted loop topology (Fig~\ref{24mm_1}d4). 
\par Upon further increase in $Re$, we observe a minute increase in the $p'_{\mathrm{rms}}$ (refer to region P in Fig~\ref{24mm_1}a). During this region, we observe that the $p'$ exhibits period 2 oscillations (Fig~\ref{24mm_1}e1). The corresponding reconstructed phase space exhibits a twisted ring topology (Fig~\ref{24mm_1}e2) and the amplitude spectrum has peaks at 1000 Hz and 2000 Hz (Fig~\ref{24mm_1}e3). As the value of $Re$ is further increased, we observe a declining of $p'_{\mathrm{rms}}$ to its minimum in the region marked by A. During this region, the system exhibits low amplitude aperiodic fluctuations (Fig~\ref{24mm_1}f1). The corresponding reconstructed phase space exhibits clustered topology (Fig~\ref{24mm_1}f2) and the amplitude spectrum is characterized by a broad band distribution (Fig~\ref{24mm_1}f3), indicating the aperiodic and low amplitude nature of the $p'$. 
\par The bifurcation from these low-amplitude aperiodic fluctuations to the states of NLC happens through a state of intermittency. Figure~\ref{24mm_1}(g1) shows the $p'$ consisting of epochs of periodic oscillations amidst epochs of aperiodic fluctuations. This switching between these apparently distinct states manifests itself as a clustered topology surrounded by a ring topology in the reconstructed phase space (Fig~\ref{24mm_1}g2). The amplitude spectrum associated with the state of intermittency exhibits a peak at 1220 Hz (Fig~\ref{24mm_1}g3), which corresponds to the frequency of oscillations of the states of NLC, before the frequency switching. That is, the system shows a gradual emergence of periodicity from aperiodicity.
\par The bifurcation from CLC to NLC follows an intricate route, namely, a transition regime characterized by high amplitude period 2 epochs amidst low amplitude NLC, period 2 oscillations, followed by a bifurcation to aperiodic fluctuations and intermittency.  When a secondary parameter of the system is varied, the governing nonlinearities of the system change \cite{etikyala2017change, qn17-x37z}, and as a consequence, the nature of the transition changes. To investigate the change in the nature of the transition in an aeroacoustic system, we vary the spacing between the orifices as a secondary parameter. 

\subsection{Canard explosion}

\begin{figure*}[h!]
\centering
\includegraphics[height=12cm]{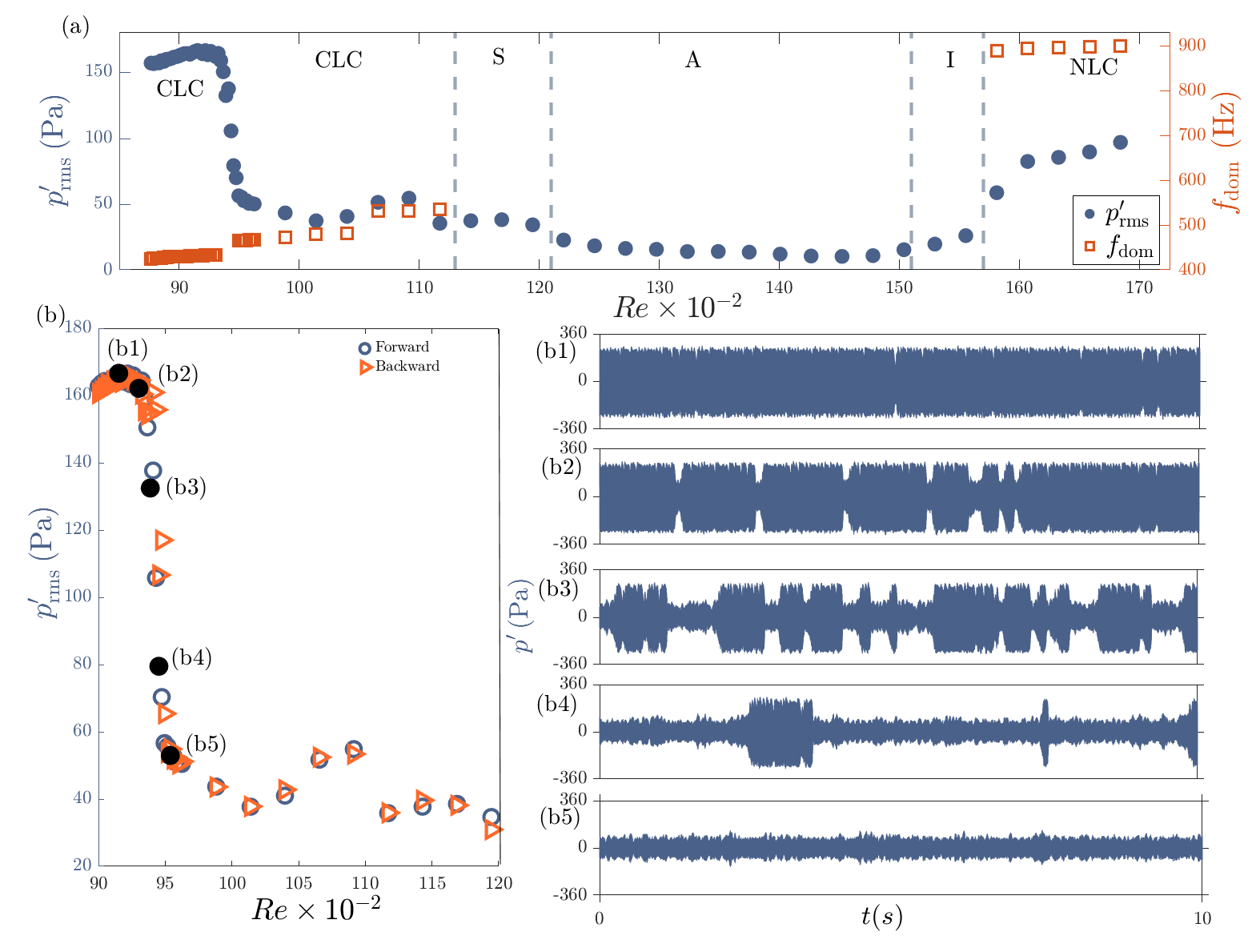}
\captionsetup{
  justification=raggedright,
  singlelinecheck=false
}
\caption{
Transition between states of LCOs consists of canard explosion (a) Variation of $p'_{\mathrm{rms}}$ and $f_{\mathrm{dom}}$ with $Re$ during the transition observed when the orifices are spaced at 18 mm apart. (b) Variation of $p'_{\mathrm{rms}}$ during the forward and backward path of $Re$ in fine steps during the canard explosion. The $p'$ corresponding to the high and low amplitude CLC, respectively (b1) and (b5). During the canard explosion, the system exhibits irregular bursting between the high and low amplitude CLCs (b2),(b3) and (b4).   The transition from CLC to NLC follows the sequence: high amplitude period 2 epochs amidst NLC (S), aperiodicity (A) and intermittency (I). 
}
\label{18mm}
\end{figure*}

At an orifice spacing of 18 mm, we observe a transition, unique for its steep declining trend in $p'_{\mathrm{rms}}$. Firstly, when the system is made to operate in the range of $Re$ of 87000 to 93000 (refer to Fig~\ref{18mm}a), we observe that the system exhibits clean LCOs. Upon an increase in $Re$, we observe that the system bifurcates to another CLC of low amplitude through a steep decline in $p'_{\mathrm{rms}}$. Furthermore, note that the shift in the $f_{\mathrm{dom}}$ associated with this bifurcation from high amplitude CLC to low amplitude CLC, unique for its steep decline in the $p'_{\mathrm{rms}}$, is from 440 Hz to 470 Hz.  When compared to the previous transition, interestingly, this shift in $f_{\mathrm{dom}}$ is similar to the observed shift in $f_{\mathrm{dom}}$, associated with the gradual decline of $p'_{\mathrm{rms}}$. To study the $p'$ during this range of $Re$ corresponding to the steep decrease in the $p'_{\mathrm{rms}}$, we vary the $Re$ at a much finer resolution possible (in steps of 22).
\par A dynamical system, when operating close to its bifurcation point, has a natural tendency to switch between its disparate states (prior to and post the bifurcation) either due to the noise inherent in the system or an unforeseen and unwanted external perturbation \cite{billings2010switching}. If this noise-driven switching is between high and low amplitude CLCs, the $p'_{\mathrm{rms}}$ of the states corresponding to this stochastic switching would lie in between the states of the corresponding high and low amplitude CLCs. The observation that the $p'_{\mathrm{rms}}$  of the states corresponding to the steep decline in the transition does indeed lie in between the  $p'_{\mathrm{rms}}$ corresponding to the high and low amplitude CLCs, suggests that it could be a manifestation of this noise driven switching. However, the monotonic decline implies the contrary, that the system follows a path, which is steep, during the bifurcation from high to low amplitude CLC. 
\par We study the backward transition when the control parameter is instead decreased, for the system may not retrace the steep path if the states during the steep decline in the transition correspond to the noise driven switching. Figure~\ref{18mm}(b) shows the forward and reverse path during the canard explosion.  We observe that the system retraces its path upon backward variation of the control parameter. This suggests that the steep decline of the $p'_{\mathrm{rms}}$ is not a consequence of mere noise sources (external or internal), but an inherent path followed by the system during the bifurcation from high amplitude CLC to low amplitude CLC.
\par Furthermore, this profile of steep decline suggests that the system is sensitive to control parameter variation. However, when this sensitivity extends to the fine fluctuations in the control parameter due to the inherent turbulent fluctuations in the system, the states corresponding to the envelope of steep decline should manifest as bursting between the disparate LCOs \cite{10.1063/5.0223320}. Figures~\ref{18mm}(b2),(b3) and (b4) shows $p'$ during the steep decline envelope (marked by black in Fig~\ref{18mm}b). Apparent in these time series of $p'$ is the irregular bursting between high and low amplitude epochs, which correspond to high and low amplitude CLCs of different frequencies, respectively. The irregularity in the bursting suggests that it is driven by the inherent turbulent fluctuations . The sensitivity of the system to the fluctuations in the control parameter is characteristic of canard explosions \cite{10.1063/5.0223320}.
\par At the value of $Re$ of 10400, similar to the previous transition, we observe a change in the trend of  $p'_{\mathrm{rms}}$ from declining to increasing. Further, associated with this change in the trend, the shift in $f_{\mathrm{dom}}$  is the same as in the previous transition, i.e., from 470 to 530. Upon further increase in $Re$, we observe a bifurcation to NLC through a sequence of dynamical states (marked as regions S, A, and I in Fig~\ref{18mm}a). Furthermore, compared to the previous transition, the number of associated dynamical states is decreased. More precisely, upon an increase in $Re$ beyond the region of CLCs, firstly, similar to the previous transition, we observe that the $p'$ exhibits brief high amplitude period 2 epochs amidst low amplitude NLC (region S in Fig~\ref{18mm}a). Following this, we observe that the system immediately bifurcates to low amplitude aperiodic fluctuations (region A in Fig~\ref{18mm}a), in contrast to the bifurcation through a period 2 state as observed in the previous transition. Subsequently, the system shows a gradual emergence of periodicity, i.e., the system undergoes bifurcation to NLC through a state of intermittency (region I in Fig~\ref{18mm}a). 
\subsection{Abrupt transition}
Hitherto in this study, we have seen different bifurcations between LCOs, namely, a canard explosion and continuous bifurcation through a sequence of dynamical states. Many dynamical systems exhibit abrupt bifurcation between different dynamical states of the system \cite{bestelmeyer2011analysis,durstewitz2010abrupt,lockwood2001abrupt,bhavi2023abrupt}. Abrupt bifurcation between qualitatively distinct states of the system has received the main focus in the literature owing to its disastrous consequences. We observe a qualitatively distinct transition, unique for its abrupt bifurcation between LCOs, when the spacing between the orifices is further decreased to $16$mm. 

\begin{figure*}[h!]
\centering
\includegraphics[height=4.5cm]{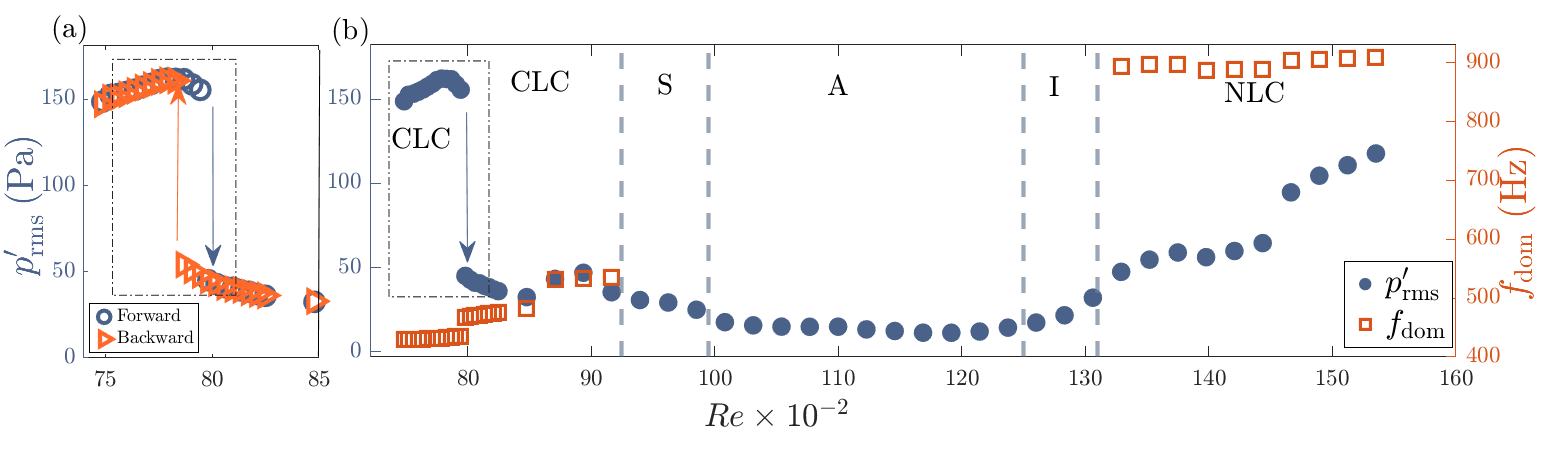}
\captionsetup{
  justification=raggedright,
  singlelinecheck=false
}
\caption{
Abrupt transition between states of LCOs. (b) Variation of $p'_{\mathrm{rms}}$ and $f_{\mathrm{dom}}$ with $Re$ during the transition observed when the orifices are spaced at $16$ mm apart. (a) We observe a hysteresis region associated with the abrupt bifurcation between the high and low amplitude CLCs. Furthermore, the bifurcation from CLC to NLC follows the same sequence as observed in the canard transition, that is, the sequence consists of high amplitude period 2 epochs amidst NLC (S), aperiodicity (A) and intermittency (I). 
}
\label{16mm}
\end{figure*}
Figure~\ref{16mm}(a) shows the transition between LCOs when the orifices are spaced at 16 mm apart. Here, the transition between high to low amplitude CLC is characterized by an abrupt fall in $p'_\mathrm{rms}$. Associated with this abrupt bifurcation, hysteresis is observed while we vary the control parameter in the backward direction. Figure~\ref{16mm}(b) shows the hysteresis zone associated with the bifurcation from CLC to CLC.  Furthermore, the frequency of oscillations of CLCs associated with this abrupt bifurcation corresponds to the CLCs associated with the canard bifurcation in the previous transition. Consequently, we note that, upon varying the orifice spacing as a secondary parameter of the system, the nature of the bifurcation between CLCs changes from a canard explosion to an abrupt bifurcation. 
\par We note that the rest of the transition exhibits close similarity to the previous transitions. That is, similar to the previous two transitions, the shift in $f_{\mathrm{dom}}$ associated with the change in the trend of $p'_{\mathrm{rms}}$ (from declining to increasing at the value of Re of 8500) is from 470 to 530.  Upon further increase in $Re$ beyond the states of CLC, we observe that the system undergoes a bifurcation to NLC through a sequence of dynamical states, similar to the previous transition. Moreover, when the orifice spacing is further decreased below 16 mm, we observe a transition, where the sequence of dynamical states associated with the bifurcation from CLC to NLC, compared to the transitions hitherto in this study, has further been simplified.

\begin{figure*}[h!]
\centering
\includegraphics[height=5cm]{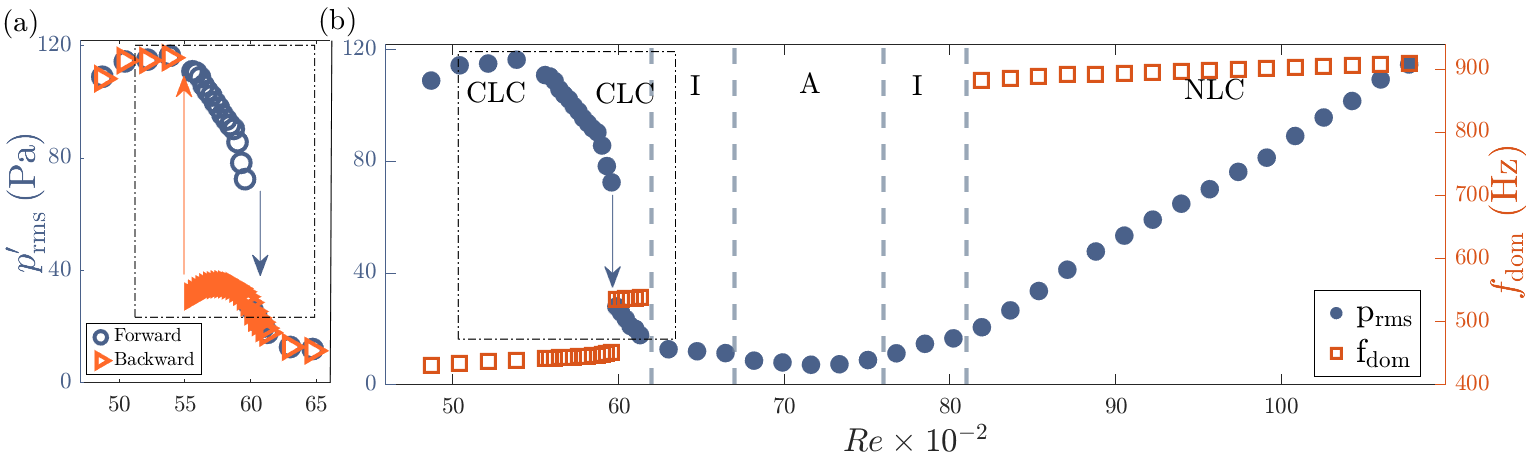}
\captionsetup{
  justification=raggedright,
  singlelinecheck=false
}
\caption{
Abrupt transition between states of LCOs. (b) Variation of $p'_{\mathrm{rms}}$ and $f_{\mathrm{dom}}$ with $Re$ during the transition observed when the orifices are spaced at 12 mm apart. (a) Hysterisis region associated with the abrupt bifurcation between the high and low amplitude CLCs. We observe that the sequence associated with the bifurcation from CLC to NLC is simplified when compared to the earlier transitions. 
This simplified sequence consists of intermittency (I), aperiodicity (A) and intermittency (I). 
}
\label{12mm}

\end{figure*}

Figure~\ref{12mm}a shows the transition when the orifices are spaced at 12 mm. As the $Re$ is increased through 6000, note the abrupt fall in the $p'_{\mathrm{rms}}$ and the sudden jump in the $f_{\mathrm{dom}}$ corresponding to the states of CLCs.  The jump in the $f_{\mathrm{dom}}$ is from 450 Hz to 535 Hz. The bistable region associated with this bifurcation is shown in Fig~\ref{12mm}(b). However, during the previous abrupt transition, the shift in $f_{\mathrm{dom}}$ associated with the abrupt bifurcation is from 440 Hz to 470 Hz. This is followed by a shift from 470 Hz to 530 Hz, associated with the change in the trend of $p'_{\mathrm{rms}}$ within the region identified as CLC. That is, in the current abrupt transition (refer to fig~\ref{12mm}a), we do not observe any CLC of intermediate frequency, either in the forward or in the backward path. In addition to this, we observe that both abrupt transitions differ in their sequence of dynamical states associated with the transition from CLC to NLC.
\par With a systematic increase in $Re$ beyond CLC, the system first gradually bifurcates to a state of aperiodic fluctuations (region A in Fig~\ref{12mm}a). Following this, upon further increase in $Re$, the system exhibits a gradual emergence of periodicity to the state of NLC. In other words, the regions I correspond to states of intermittency, characterized by the presence of epochs of aperiodic fluctuations and periodic oscillations. 
\par In summary, upon decreasing the distance between the orifices as a secondary parameter, the nature of the transition between CLC and NLC changes from continuous to abrupt through a canard explosion. Associated with this change, we show that the dynamical states characterizing the route to transition between CLC and NLC are altered.  The continuous transition, occurring at large spacing between the orifices, exhibits an intricate route, consisting of different dynamical states and transient regimes. On the other hand, one of the abrupt transitions occurring at smaller distances between the orifices exhibit a route that delineates desynchronization to aperiodic fluctuations through intermittency, followed by transition to instability through intermittency.  Further,  we identify two distinct abrupt transitions, differing in their route to transition between the LCOs.

\section{Conclusion and Discussion}
In our laboratory-scale experimental setup consisting of ducted flow past two orifices, we vary the Reynolds number as a primary parameter to observe the transition between the states of clean LCO and noisy LCO. During this transition, the system exhibits various dynamical states such as aperiodicity, period 2 oscillations, intermittency and a transition regime characterized by switching between period 2 and period 1 oscillations. Upon a decrease in the orifice spacing as a secondary parameter, we observe that the nature of the transition changes from continuous to abrupt.
\par Several dynamical systems spanning ecology, magneto-hydrodynamics, quantum mechanics, and thermoacoustics exhibit a change in the nature of the transition from continuous to abrupt upon variation in a secondary parameter. \citet{doi:10.1126/sciadv.abe3824}, in their recent mathematical investigation consisting of introducing higher order nonlinearities through secondary parameter variation in three dynamical systems, have shown this change to be a universal phenomenon. When compared to this traditional notion of change in the nature of transition from continuous to discontinuous, intriguingly, we observe that aeroacoustic flows exhibit an intricate change. We discover a metamorphosis of transition between LCOs in aeroacoustic instabilities. We show that aeroacoustic flows exhibit a change in the nature of the transition between LCOs from continuous to abrupt via a canard explosion. Furthermore, we show a change in the sequence of dynamical states associated with this change in the nature of the transition. 
\par While abrupt bifurcations show hysteresis upon backward variation of the primary parameter, continuous bifurcations retrace their path. Canard explosions, however, are dissimilar to the abrupt bifurcations, for they retrace their path upon backward variation of the primary parameter. Furthermore, due to the sensitivity to the variations in the primary parameter, canard explosions exhibit a continuous yet rapid change in the $p'_{\mathrm{rms}}$. However, the states of bursting between the desperate dynamical states (high and low amplitude CLC) during the canard explosion, driven by the turbulent fluctuations in the $Re$, distinguish it from the continuous bifurcation.  
\par Turbulence is a multiscale phenomenon \cite{kolmogorov1941local}. It enhances local interactions between subsystems across spatial locations and scales ranging from the Kolmogorov scales to the integral scales.  By decreasing the distance between the orifices,  the turbulence within the shear layer is further amplified due to the increased upstream influence of the second orifice.  Further, increasing the mass flow rate increases the intensity of turbulence and local-scale interactions across subsystems.  This phenomenon could be a possible governing mechanism for the metamorphosis of transition.
\par In conclusion, aeroacoustic instabilities, driven by the inherent turbulence, exhibit a metamorphosis of transition between LCOs from continuous to discontinuous through a canard explosion. While many dynamical systems exhibit both continuous and discontinuous transitions, the evolution of the transition from one to another is largely unexplored.  Our study unravels a change in the nature of the transition from continuous to discontinuous via a canard explosion. Understanding the metamorphosis of transitions in a dynamical system can aid in the development of low-cost control and preventive strategies. In this context, the present study enables the fine-tuning of parameters such as the spacing between orifices in applications such as solid rocket motors and industrial pipeline systems. Furthermore, developing lower-order models for the metamorphosis of a transition would enhance our understanding of the governing nonlinearities of the system and similarities between the different systems. However, capturing the full scope and intricacies of the metamorphosis of a transition remains a challenging task. 
\par In addition, further spatiotemporal measurements of the velocity field confined between the two orifices would provide valuable information on (i)  the change in the dynamics of the hydrodynamic field leading to the identified transitions and (ii) information on the synchronization between the acoustic and hydrodynamic subsystems associated with the metamorphosis.

\bibliography{aipsamp2}
\nocite{*}
\end{document}